\documentclass{PoS}

\title{The Role of Provenance Management in Accelerating the Rate of Astronomical Research}

\ShortTitle{Provenance Management in Astronomy}

\author{\speaker{G. Bruce Berriman}\\
        NASA Exoplanet Science Institute, Infrared Processing and Analysis Center, California Institute of Technology
770 South Wilson Avenue, Pasadena, CA  91125, USA
\\
        E-mail: \email{gbb@ipac.caltech.edu}}

\author{{Ewa Deelman}\\
       Information Sciences Institute, University of Southern California,
4676 Admiralty Way, Suite 1001, Marina del Rey, CA 90292, USA
\\
        E-mail: \email{deelman@isi.edu}}

\abstract{The availability of vast quantities of data through electronic archives has transformed astronomical research. It has also enabled the creation of new products, models and simulations, often from distributed input data and models, that are themselves made electronically available. These products will only provide maximal long-term value to astronomers when accompanied by records of their provenance; that is, records of the data and processes used in the creation of such products. We use the creation of image mosaics with the Montage grid-enabled mosaic engine to emphasize the necessity of provenance management and to understand the science requirements that higher-level products impose on provenance management technologies. We describe experiments with one technology, the "Provenance Aware Service Oriented Architecture" (PASOA), that stores provenance information at each step in the computation of a mosaic. The results inform the technical specifications of provenance management systems, including the need for extensible systems built on common standards. Finally, we describe examples of provenance management technology emerging from the fields of geophysics and oceanography that have applicability to astronomy applications
}

\FullConference{Accelerating the Rate of Astronomical Discovery, sps5\\
		August 11-14, 2009\\
		Rio de Janeiro, Brazil}

\begin{document}

\section{ Introduction}

Astronomers need to understand the technical content of data sets and evaluate published claims based on them. All data products and records from all the steps used to create science data sets ideally would be archived, but  the volume of data would be prohibitively high. The high-cadence surveys currently under development will exacerbate this problem; the Large Synoptic Survey Telescope alone is expected to deliver 60 PB of just raw data in its operational lifetime.  There is therefore a need to create records of how data were derived $-$ provenance - that contain sufficient information to enable replication of the data.  A report issued by the National Academy of Sciences dedicated to the integrity of digital data recommends the curation of the provenance of data sets as part of its key recommendations \cite {NA09}.

Provenance records must meet strict specifications if they are to have value in supporting research. They must capture the algorithms, software versions, parameters, input data sets, hardware components and computing environments.  The records should be standardized and captured in a permanent store that can be queried by end users. 
In this paper, we describe how the Montage image mosaic engine acts as a driver for the application in astronomy of provenance management methodologies now in development.   Provenance management is an active field in many areas of science, and we describe work in earth sciences and oceanography that has applicability to astronomy.  \cite{Deelman09} describes provenance management in more detail. 

\section{ Montage : A Case Study for Provenance Management}
\subsection {What is Montage?}
Montage (http://montage.ipac.caltech.edu) is a toolkit for aggregating astronomical images in Flexible Image Transport System (FITS) format into mosaics. Its scientific value derives from three features of its design:
\begin{itemize}
\item It uses algorithms that preserve the calibration and positional (astrometric) fidelity of the input images to deliver mosaics that meet user-specified parameters of projection, coordinates, and spatial scale. It supports all projections and coordinate systems in use in astronomy.
\item It contains independent modules for analyzing the geometry of images on the sky, and for creating and managing mosaics.
\item It is written in American National Standards Institute (ANSI)-compliant C, and is portable and scaleable $Ð$ the same engine runs on desktop, cluster, supercomputer environments or clouds running common Unix-based operating systems. 
\end{itemize}

There are four steps in the production of an image mosaic:

1. Discover the geometry of the input images on the sky from the input FITS keywords and use it to calculate the geometry of the output mosaic on the sky.

2. Re-project the input images to the spatial scale, coordinate system, World Coordinate System (WCS)- projection, and image rotation.

3. Model the background radiation in the input images to achieve common flux scales and background level across the mosaic.

4. Co-add the re-projected, background-corrected images into a mosaic.

Each production step has been coded as an independent engine run from an executive script. Figure 1 illustrates the second through fourth steps for the simple case of generating a mosaic from three input mosaics.  In practice, as many input images as necessary can be processed in parallel, limited only by the available hardware.

\begin{figure}
\includegraphics [width=1.1\textwidth] {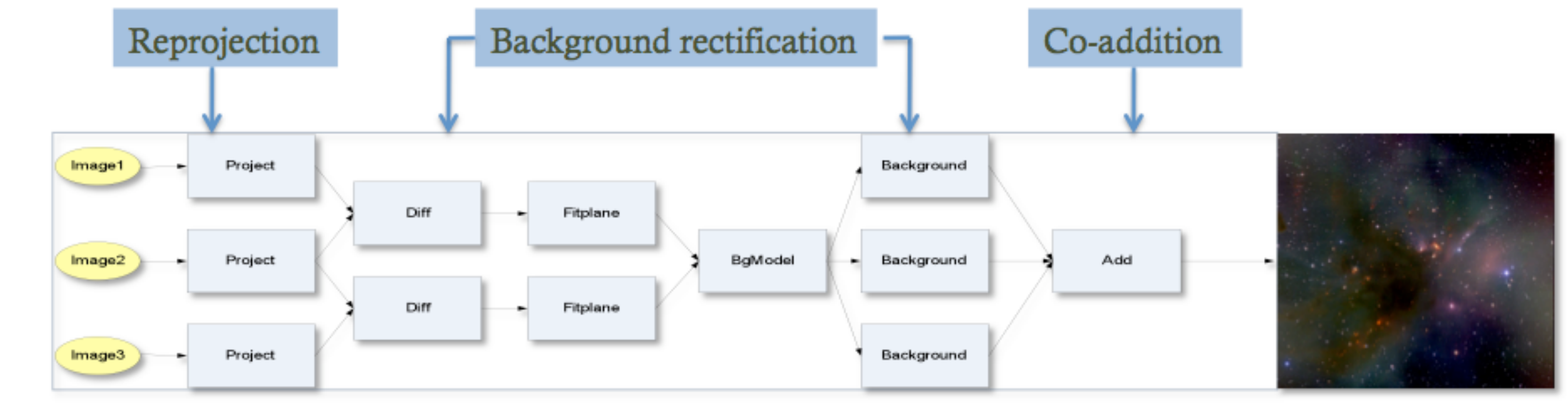}
\caption{The processing steps used in computing an image mosaic with the Montage engine.}
\label{Fig 1}
\end{figure} 

\subsection{ Production of Mosaics}
In the production steps shown in Figure 1, the files output by one step become the input to the subsequent step. That is, the reprojected images are used as input to the background rectification. This rectification itself  consists of several steps that fit a model to the differences between flux levels  of each image, and in turn the rectified, reprocessed images are input to the co-addition engine. Thus the production of an image mosaic actually generates a volume of data that is substantially greater than the volume of the mosaic. Table 1 illustrates this result for two use cases that return 3-color mosaics from the Two Micron All Sky Survey (2MASS) images (see http://www.ipac.caltech.edu/2mass/releases/allsky/doc/explsup.html). One is a 6 deg sq mosaic of $\rho$ Oph and the second is an All Sky mosaic.  The table makes clear that the volume of intermediate products exceeds the mosaic size by factors of 30 to 50. The Infrared Processing and Analysis Center (IPAC) hosts an on-request image mosaic service (see Section 3)  that delivers mosaics of user-specified regions of the sky, and it currently receives 25,000 queries per year. Were mosaics of the size of the  $\rho$ Oph mosaic processed with such frequency, the service would produce 3.8 PB of data each year. Such volumes are clearly too high to archive.

\begin{table}[ht]
\centering 
\begin{tabular}{l r r } 
\hline\hline 
 & $\rho$ Oph 6 deg sq & All Sky Mosaic   \\ [0.5ex] 
\hline 
 \# input images & 4,332 &   4,121,439 \\ 
\# comp. steps & 25,258 & 24,030,310 \\
\# intermediate products & 67,300  &  61,924,260\\
 Size of intermediate products & 153 GB & 126 TB   \\
  Mosaic Size  & 2.4 GB & 4 TB \\ 
  Annual Volume  & 3.8 PB  &  ... \\ [1ex] 
\hline 

\end{tabular}
\caption{Estimates of Files Generated in the Production of Image Mosaics. See text for an explanation of {\it Annual Volume.} } 
\label{table:nonlin} 
\end{table}

\subsection{The Scientific Need To Reprocess Mosaics}

Montage makes three assumptions and approximations that affect the quality of the mosaics:

\begin{itemize}

\item  Reprojection involves redistributing the flux from the input pixel pattern to the output pixel pattern.  Montage uses a fast, custom algorithm that approximates tangent plane projections\footnote [1] {  Geometric projections of the celestial sphere onto a tangent plane from a center of projection at the center of the sphere}  as polynomial approximations to the pixel pattern on the sky, which can produce small distortions in the pixel pattern of the mosaic. 

\item There is no physical model of the sky background that predicts its flux as a function of time and wavelength. Montage assumes that the sky background is only significant at the lowest spatial frequencies, and rectifies the flux at these frequencies to a common level across all the input images. This approximation can confuse background flux with an astrophysical source present at the same frequencies, such as extended diffuse emission in a nebula or dust cloud.

\item Co-additions of the reoprojected, rectified images are weighted to not take into account outliers due to e.g. residual cosmic ray hits.

\end{itemize}

Users have two options in investigating the impact of these three factors, and both involve knowing the provenance of the mosaics:
1. Analyze the output from intermediate steps to understand how the features in the mosaic originate.
2. Replace modules with implementations of new algorithms, such as a custom background rectification, and reprocess the  mosaic.

\section{Information Needed In Provenance Records}

Column 1 of Table 2 lists all the information needed to specify a provenance record for an image mosaic.  To illustrate the current quality of provenance recording, column 2 describes the provenance information that is made available to users by an on-line, on-request image mosaic service at http://hachi.ipac.caltech.edu:8080/montage/. This service is hosted at IPAC, and returns mosaics of 2MASS, Sloan Digital Sky Survey (SDSS) and Digitized Sky Surveys at Space Telescope (DSS) images. When processing is complete, users are directed to a web page that contains links to the mosaic and to processing information. It is to the contents of these pages that column 2, table 2 refers. 

The only information that is permanently recorded are the runtime parameters that specify the properties of the image mosaic $-$the coordinate system, projection, spatial sampling and so on $-$ written as keywords in the mosaic file header.  The file itself, as well as log files and the traceability to the input images, are deleted after 72 hours (but these can be reproduced if the user has a record of the specifications of the mosaic requested).  There is no record of the execution environment, and the algorithm and software information are described in the project web page, and presume that users know where to find them and that the web pages do not become stale.

\begin{table}[ht]
\centering 
\begin{tabular}{l l} 
\hline\hline 
{\bf Information} & {\bf Recorded In On-Request Service?} \\ [0.5ex] 
\hline 
{\bf Algorithms} &   \\ 
Algorithm Design Documents & Accessible from Montage web page \\
Algorithm Version & Accessible from Montage web page \\
{\bf Execution Environment} & \\
Specific hardware & No \\
OS and version & 	No \\
Process Control and Management Tools & No \\
{\bf Software} &  \\
Software Source Code, version	& Accessible from Montage web page \\
Software Build Environment, version &  Accessible from Montage web page  \\
Compiler, version & Accessible from Montage web page  \\
Dependencies and versions & Accessible from Montage web page  \\
Test Plan Results & Accessible from Montage web page  \\
{\bf Runtime} &  \\
Parameters	& Included in output files \\
Input files, version	& Retained for 72 hours after completion of job \\
Output Files, Log Files &	Retained for 72 hours after completion of job \\ [1ex] 
\hline 
\end{tabular}
\caption{Comparison of Required and Recorded Provenance Information} 
\label{table:nonlin} 
\end{table}

\section{Experiments in Recording Provenance Information}
The previous section reveals an obviously unsatisfactory state of affairs.   We have therefore investigated how astronomers may take advantage of methodologies already under development in other fields to create and manage a permanent store of provenance records for the Montage engine.  When complete, these investigations are intended to deliver an operational provenance system that will enable replication of any mosaic produced by Montage.

\subsection{Characteristics of Applications and Provenance Management}

The design of Montage is  well suited for the creation of provenance records, as follows (see  \cite {Groth09}  for more details):

\begin{itemize}
\item It is deterministic; that is, processing a common set of input files will yield the same output mosaic.
\item It is component based, rather than monolithic.
\item It is self-contained and requires, e.g., no distributed services.
\item It runs on all common hardware platforms.
\item It inputs data in self-describing standard formats. 
\item Its input data are curated and served over the long term.
\end {itemize}

\subsection{Capturing the Provenance of Montage Processing}

Many provenance systems are embedded in processing environments, which offer the benefits of efficient collection of self-contained provenance records, but at the cost of ease of interoperation with other provenance systems.  Given that Montage can be run as a pipeline, it too can employ such a system,  and indeed (\cite {Groth09}) has demonstrated this.  In this paper, we will report instead on efforts to leverage an existing methodology to create a standardized provenance store that can interoperate with other applications.  The methodology is the {\it Provenance Aware Service Oriented Architecture} (PASOA) (\cite{Miles07}), an open source architecture already used in fields such as aerospace engineering, organ transplant management, and bioinformatics. In brief, when applications are executed they produce documentation of the process that is recorded in a {\it provenance store},  essentially a repository of provenance  documents and records. The store is housed in a database so that provenance information can be queried and accessed by other applications .

In our investigation, Montage was run with the Pegasus framework \cite{Deelman05}.   Pegasus was developed to map complex scientific workflows onto distributed
resources. It operates by taking the description of the processing flow in Montage (the abstract workflow) and mapping it onto the physical resources that will run it, and records this information in its logs. It allows Montage to run on multiple environments and takes full advantage of the parallelization inherent in the design. Pegasus has been augmented with PASOA to create a provenance record for Montage in eXtended Markup Language (XML) that captures the information identified in Table 2.  We show a section of this XML structure below, captured during the creation of a mosaic of M17:

\tiny
\begin{verbatim}

<?xml version="1.0" encoding="ISO-8859-1"?>
<invocation xmlns="http://vds.isi.edu/invocation" xmlns:xsi="http://www.w3.org/2001/XMLSchema-instance" 
xsi:schemaLocation="http://vds.isi.edu/invocation http://vds.isi.edu/schemas/iv-1.10.xsd" version="1.10" 
start="2007-03-26T16:25:54.837-07:00" duration="12.851" transformation="mShrink:3.0" derivation="mShrink1:1.0" 
resource="isi_skynet" hostaddr="128.9.233.25" hostname="skynet-15.isi.edu" pid="31747" uid="1007" user="vahi" 
          gid="1094" 
group="cgt" umask="0022">
<prejob start="2007-03-26T16:25:54.849-07:00" duration="5.198" pid="31748">
  <usage utime="0.010" stime="0.030" minflt="948" majflt="0" nswap="0" nsignals="0" nvcsw="688" nivcsw="4"/>
  <status raw="0"><regular exitcode="0"/></status>
  <statcall error="0">
    <!-- deferred flag: 0 -->
    <file name="/nfs/home/vahi/SOFTWARE/space_usage">23212F62696E2F73680A686561646572</file>
    <statinfo mode="0100755" size="118" inode="20303958" nlink="1" blksize="32768" blocks="8" 
          mtime="2007-03-26T11:06:24-07:00" 
atime="2007-03-26T16:25:52-07:00" ctime="2007-03-26T11:08:12-07:00" uid="1007" user="vahi" gid="1094" group="cgt"/>
  </statcall>
  <argument-vector>
    <arg nr="1">PREJOB</arg>
  </argument-vector>
</prejob>
<mainjob start="2007-03-26T16:26:00.046-07:00" duration="2.452" pid="31752">
  <usage utime="1.320" stime="0.430" minflt="496" majflt="8" nswap="0" nsignals="0" nvcsw="439" nivcsw="12"/>
  <status raw="0"><regular exitcode="0"/></status>
  <statcall error="0">
    <!-- deferred flag: 0 -->
    <file name="/nfs/home/mei/montage/default/bin/mShrink">7F454C46010101000000000000000000</file>
    <statinfo mode="0100755" size="1520031" inode="1900596" nlink="1" blksize="32768" blocks="2984" 
mtime="2006-03-22T12:03:36-08:00" atime="2007-03-26T14:16:36-07:00" ctime="2007-01-11T15:13:06-08:00"
 uid="1008" user="mei" gid="1008" group="mei"/>
  </statcall>
  <argument-vector>
    <arg nr="1">M17_1_j_M17_1_j.fits</arg>
    <arg nr="2">shrunken_M17_1_j_M17_1_j.fits</arg>
    <arg nr="3">5</arg>
  </argument-vector>
</mainjob>
\end{verbatim}

\normalsize

Our experiments with PASOA have been successful and a next step will be to deploy it as part of operational system. 

\section{Applications in Earth Sciences and Oceanography}

While the work described above is an advanced experimental stage,  Earth Sciences and Oceanography projects have for a number of years exploited operational provenance management systems  {\cite{Deelman09}).  We would suggest that astronomy has much to learn from these projects. Here we describe two examples, one involving an integrated pipeline, and one involving a complex data system that uses many instruments collecting a complex and dynamic data set.

\subsection{Example 1: The Moderate Resolution Imaging Spectroradiometer (MODIS)}

An instrument launched in 1999 aboard the Terra platform, MODIS scans the Earth in 36 bands every two days.  The raw ("level 0") data are transformed into calibrated, geolocated prodcuts  ("level 1B"), which are then aggregated into global data products ("level 2") that are the primary science products. Examples include a global vegetative index map and a global sea surface temperature map.  The raw data are archived permanently, but  the level 1B data are much too large to archive. These data are retained for 30-60 days only.  Consequently,  the MODIS archive records all the process documentation needed to reproduce the Level 1B data from the raw satellite data. The process documentation includes the algorithms used, their versions, the original source code, a complete description of the processing environment and even the algorithm design documents themselves \cite{Tilmes08}.

\subsection{Example Two: The Monterey Bay Aquarium Shore Side Data System (SSDS)}

For the past four years, the SSDS has been used to track the provenance of complex data sets form many sources \cite{McCann08}.  Oceanographers undertake  campaigns that involve taking data from multiple sources $-$ buoys, aircraft, underwater sensors, radiosondes and so on. These instruments measure quantities such as salinity and amount of chlorophyll.  These data  are combined with published data including satellite imagery in simulations to predict oceanographic features, such as seasonal variations in water levels.   The SDSS was developed to track the provenance of the data measured in the campaigns in standardized central repository. Scientists use SDSS to track back from derived data products to the metadata of the sensors including their physical location, instrument and platform. The system  automatically populates metadata fields, such as the positions of instruments on moving platforms.

\section{Conclusions}

\begin{itemize}
\item Tracking the provenance of data products will assume ever-growing importance as more and larger data sets are made available to astronomers.
\item Methodologies such as PASOA are in use in aerospace and bioinfomatics applications and show great promise for providing provenance stores for astronomy.
\item Earth Science projects routinely track provenance information. There is much that astronomy can learn from them.
\item There is also an effort in the provenance community to standardize on a provenance model \cite{opm}, intended to foster interoperability between provenance systems and spur on the development of generic provenance capture and query tools.
\end{itemize}

\end{document}